\documentclass[aps,prb,twocolumn,superscriptaddress,amsmath,amssymb,citeautoscript]{revtex4-2}

\usepackage{graphicx}
\usepackage{dcolumn}
\usepackage{booktabs}
\usepackage{bm}
\usepackage{color}
\usepackage{multirow}
\usepackage{ulem}
\usepackage{xcolor}
\usepackage{hyperref}
\hypersetup{colorlinks = {true},
    		citecolor  = {blue},
    		urlcolor   = {blue},
    		linkcolor  = {blue}}

\newcommand{\Tc}{\ensuremath{T_{\rm c}}}
\newcommand{\gn}{\ensuremath{\gamma_{\rm n}}}

\newcommand{\sh}{\ensuremath{c_{p}}}
\newcommand{\cel}{\ensuremath{c_{\rm el}}}

\newcommand{\EF}{\ensuremath{E_{\rm F}}}
\newcommand{\kB}{\ensuremath{k_{\rm B}}}

\newcommand{\rxx}{\ensuremath{\rho_{xx}}}

\newcommand{\lamep}{\ensuremath{\lambda_{\rm el-ph}}}

\newcommand{\omD}{\ensuremath{\omega_{\rm D}}}

\newcommand{\IT}{In$_{x}$Te}

\newcommand{\Sup}{Supplemental Materials (SM)}

\begin{document}

\title{Enhancement of superconductivity and its relation to lattice expansion in InTe}

\author{M.~Kriener}
\email[corresponding author: ]{markus.kriener@riken.jp}
\affiliation{RIKEN Center for Emergent Matter Science (CEMS), Wako 351-0198, Japan}
\author{M.~S.~Bahramy}
\email[corresponding author: ]{m.saeed.bahramy@manchester.ac.uk}
\affiliation{Department of Physics and Astronomy, The University of Manchester, Oxford Road, Manchester M13 9PL, United Kingdom}
\author{Y.~Tokura}
\affiliation{RIKEN Center for Emergent Matter Science (CEMS), Wako 351-0198, Japan}
\affiliation{Department of Applied Physics and Quantum-Phase Electronics Center (QPEC), University of Tokyo, Tokyo 113-8656, Japan}
\affiliation{Tokyo College, University of Tokyo, Tokyo 113-8656, Japan}
\author{Y.~Taguchi}
\affiliation{RIKEN Center for Emergent Matter Science (CEMS), Wako 351-0198, Japan}
\date{\today}

\begin{abstract}
The quest to govern the driving forces behind superconductivity and gain control over the superconducting transition temperature \Tc\ is as old as the phenomenon itself. Microscopically, this requires a proper understanding of the evolution of electron-lattice interactions in their parameter space. We report such a controlled study on \Tc\ in \IT\ via fine-tuning the In stoichiometry $x$. We find that increasing $x$ from 0.84 to 1 results in an enhancement of \Tc\ from 1.3~K to 3.5~K accompanied by an increase of the electron-phonon coupling constant from 0.45 to 0.63. Employing first-principles calculations, we show that this behavior is driven by two factors, each taking the dominant role depending on $x$. For $x\lesssim 0.92$, the major role is played by the density of electronic states at the Fermi level. Above $x\sim 0.92$, the change in the density of states flattens while the enhancement of \Tc\ continues. We attribute this to a systematic softening of lattice vibrations, amplifying the electron-phonon coupling, and hence, \Tc. 

\end{abstract}

\maketitle

\section{Introduction}

Superconductivity has been the subject of continuous research since its discovery more than a century ago in 1911 by Kammerling-Onnes \cite{onnes11a}. Despite its long history, the question of what governs and how to control the superconducting transition temperature \Tc\ is still a tempting issue. Even in conventional Bardeen-Cooper-Schrieffer (BCS) superconductors, while seemingly well understood theoretically, there are various strategies how to enhance \Tc\ discussed in the literature. According to the BCS theory \cite{bardeen57a}, larger \Tc\ values are expected for enhanced phonon frequencies involved in Cooper pairing, an enhanced density of states (DOS) at the Fermi level \EF, and an increased electron-phonon interaction. Soon after, Anderson pointed out that \Tc\ in conventional superconductors is barely affected by nonmagnetic impurity scattering and weak disorder \cite{anderson59b}. Nevertheless, there are theoretical works where such imperfections are discussed to bear the potential to enhance \Tc\ \cite{belitz87a,imartin97a}. Also, the valence-skipping feature of some elements is considered to be capable of improving the superconducting pairing interaction via the so-called negative-$U$ mechanism, as pointed out by Varma \cite{varma88a,hase16a}. Doping is one common approach to exploit these mechanisms, which all work within a BCS framework. Another versatile way to control \Tc\ is by applying physical pressure $p$, which may enhance or suppress \Tc, depending on the particular system. Physical pressure is an experimental tool to manipulate solely the unit-cell volume without introducing disorder into the system, which was already phenomenologically discussed by Matthias and others in the 1950s \cite{matthias53c,roegener52a}: Arguably, an expanded lattice is often in favor of yielding larger \Tc\ values as compared to compressed lattices. 

Against this background, cubic InTe is an interesting superconductor: In is one of these valence-skipping elements found in the periodic table. At first glance, it should take its divalent state here, given the strong electron affinity of Te which tends to be in its $2-$ state. However, In$^{2+}$ is energetically unstable \cite{geller64a,hase16a,kriener18a,kriener20a} and usually takes its $1+$ or $3+$ state. A direct correlation between changes in \Tc\ and the In content $x$ in \IT\ was reported in the past but only phenomenologically explained in terms of a changing ratio of In$^{1+}$ and In$^{3+}$ ions with $x$ \cite{geller64a,geller64b}. As reported recently \cite{kriener20a}, the overall In valence state is likely to be close to $1+$ in InTe.

\begin{figure}[t]
\centering
\includegraphics[width=1\linewidth]{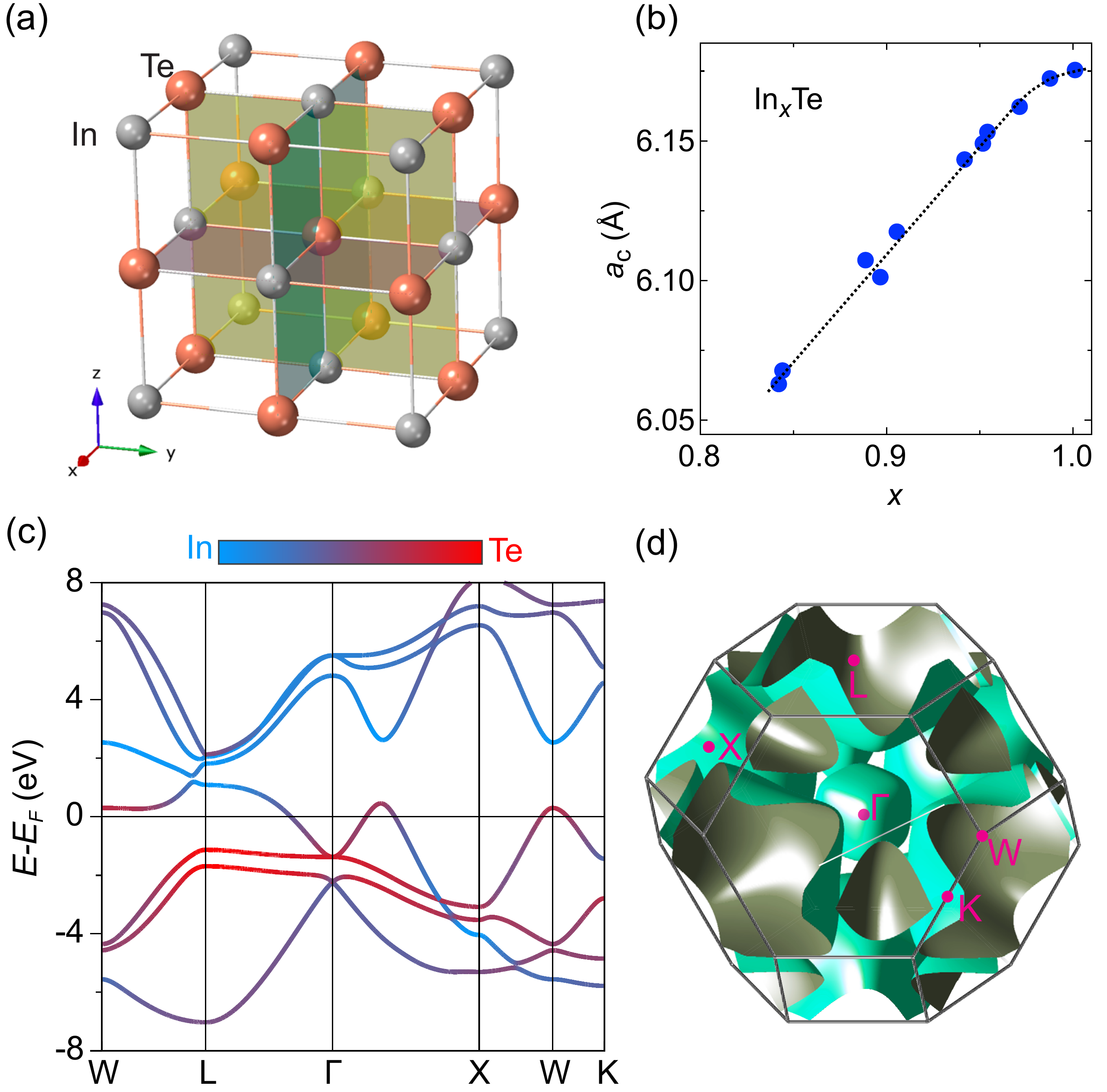}
\caption{(a) Sketch of the cubic InTe structure. (b) Cubic lattice constant $a_c$ as a function of the In content $x$. The dotted line is a guide to the eyes. (c) Calculated band structure of InTe. The colors represent the weight of the constituting orbitals of In (blue) and Te (red). (d) Brillouin zone of InTe. The positions of several high-symmetry points are labeled. There are electron- (centered at the $\Gamma$ point) and hole-like pockets ($L$ points) in the vicinity of the Fermi energy \EF.}
\label{fig1}
\end{figure}

In this paper, to elucidate the microscopic mechanism that governs \Tc, we vary the In concentration $0.84 \leq x \leq 1$ in \IT. This enables a very fine-tuned and systematic control of \Tc\ from around 1.3~K to about 3.5~K. This composition control allows to study the interrelation of DOS, phonon frequencies, electron-phonon interaction, and lattice expansion by simply changing $x$. Here, we observe an interesting crossover in the nature of the superconductivity around $x\sim 0.92$ where \Tc\ has increased to about 2.2~K: Initially this enhancement can be traced back to a concomitant increase of DOS with $x$ as it is often seen in conventional superconductors. However, above $x \gtrsim 0.92$, the DOS enhancement becomes gradual and, hence, solely DOS-based arguments cannot explain the observed continued enhancement of \Tc\ up to $x = 1$. The only material parameter which keeps changing is the cubic InTe lattice constant $a_c$, as presented in Fig.~\ref{fig1}(b). This suggests that the lattice expansion is dominant for the \Tc\ enhancement for $x\gtrsim 0.92$ in \IT. For this In concentration range $0.92 < x \leq 1$, we develop a simple model with the cubic lattice parameter as only variable. It successfully reproduces qualitatively and quantitatively the experimental evolution of DOS, electron-phonon coupling constant \lamep, and \Tc\ with $x$. A lattice-expansion-induced enhancement of the electron-phonon coupling strength is proposed to be responsible for the observed enhancement of \Tc\ in this $x$ range.

This paper is organized as follows: After summarizing the experimental and computational methods in the next Section~\ref{methods}, we present experimental results in Section~\ref{results}. Our theoretical model is introduced in Section~\ref{theory}. In Section~\ref{discussion}, theoretical and experimental results for $x>0.92$ are compared and we discuss possible origins of the continuous enhancement of \Tc\ and its fine-tuned controllability via the composition ratio. We conclude with summarizing this work. Additional data and discussions are provided in the accompanying \Sup\ \cite{Suppl}.

\section{Experimental and Computational Methods}
\label{methods}
\IT\ batches with $0.84 \leq x \leq 1$ were grown by melting stoichiometric amounts of In and Te shots in evacuated quartz glass tubes at 950$^{\circ}$C for 24 -- 48~h, and subsequently quenching into water. The resulting tetragonal InTe material was ground again and approximately 400~mg powder of each batch were used for high-pressure synthesis (5 GPa, 600$^{\circ}$C, 1~h) to obtain the metastable superconducting phase of InTe with cubic structure at ambient conditions. 

X-ray diffraction patterns were taken on these batches with an in-house x-ray diffractometer (Rigaku). All batches with $x> 0.85$ were found to be single-phase cubic InTe (space group 225; $Fm\bar{3}m$) with sharp reflection peaks, see Figs.~S1 and S2 in \cite{Suppl}. Estimated cubic lattice constants $a_c$ are plotted against $x$ in Fig.~\ref{fig1}(b), exhibiting a linear variation with a saturation tendency when approaching $x=1$. Only for the lowest In concentration $x = 0.84$ very tiny impurity peaks are seen, possibly indicating the start of the formation of other phases. To further check this, we also made one test specimen with $x = 0.79$, where these and additional impurity peaks are more pronounced. This apparently indicates the lower border of stability of cubic \IT, which seems also reflected in a slight broadening of the peak widths, cf.\ Section~S1 in \cite{Suppl} for a brief discussion. One might also suspect that reducing the In content in \IT\ causes antisite defects. However, a careful analysis of our XRD data shows that even for large In deficiency, Te remains on its regular lattice sites without forming antisite defects down to $x=0.84$, see Fig.~S3 in \cite{Suppl}.

The In concentration of all batches was checked by inductively coupled plasma atomic-emission spectroscopy (ICP-AES) chemical analyses. The results are close to the nominal values, cf.\ Section~S2 in \cite{Suppl}. Throughout the paper, the ICP results are used when referring to samples.

The superconducting critical temperatures \Tc\ of all samples were determined by temperature-dependent magnetization $M(T)$ measurements (magnetic property measurement system MPMS3 equipped with a $^3$He insert, Quantum Design). Data were taken upon heating in $B=10$~G after zero-field cooling to the base temperature. All samples exhibit sharp single superconducting transitions. The shielding fractions were roughly corrected for the demagnetization effect according to Ref.~\cite{brandt99a}. Our data suggest large superconducting phase fractions close to 100\%. In magnetization measurements, \Tc\ is defined as the intersection of a linear extrapolation of the transition in $M(T)$ with the normal-state signal, cf.\ Section~S3 in \cite{Suppl}.

Resistivity \rxx\ and specific heat \sh\ were measured on selected samples by a standard four-probe technique and relaxation method, respectively (physical property measurement system PPMS equipped with a $^3$He insert, Quantum Design). In these measurements, \Tc\ is either defined as the temperature at which the resistivity drops to zero, or as midpoint of the superconducting transition in the electronic specific heat divided by temperature $\cel/T$, cf.\ Section~S4 in \cite{Suppl}. Measurements of the resistivity under hydrostatic pressure up to approximately 2.25~GPa were performed with a clamp-type pressure cell mounted to a PPMS sample puck (pressure cell HPC-33, ElectroLab Corporation, pressure medium: Daphne 7373 oil). The  applied pressure was determined from the suppression of the superconducting \Tc\ of a simultaneously measured Pb standard sample.

The electronic and vibrational properties of InTe were calculated within density functional theory \cite{ponce16a} using Perdew-Burke-Ernzerhof exchange-correlation functional~\cite{perdew96a} and ultrasoft pseudo-potentials as implemented in the Quantum Espresso program package~\cite{giannozzi09amod,giannozzi17amod,qepp}. The plane-wave cut-off energy was set to 35~Ry. The relativistic effects, including spin-orbit coupling, were fully considered. An fcc lattice with a variable lattice constant was chosen for InTe. The corresponding Brillouin zone (BZ) was sampled by a 24$\times$24$\times$24 $k$-mesh. 
The phonon modes, Eliashberg spectral function $\alpha^2F(\omega)$, and electron-phonon coupling constant \lamep\ were computed using density-functional perturbation theory employing a $4\times4\times4$~$q$-mesh. The calculation of the superconducting transition temperature \Tc\ was done by means of Mc-Millan's equation modified by Allen and Dynes~\cite{mcmillan68a,allen75a} using the screened Coulomb potential $\mu^{*} = 0.1$.

\section{Results}
\label{results}
A schematic plot of the face-centered cubic structure of superconducting InTe is shown in Fig.~\ref{fig1}(a) \cite{banus63a,boemmel63a}. Each Te is octahedrally coordinated with six In ions as its first-nearest neighbors (1NN). The second-nearest neighbors are twelve Te ions, forming cuboctahedral coordination. Te as an anion needs two electrons to complete its $5p$ shell. The In cations can afford to transfer one electron to their 1NN \cite{kriener20a}. To satisfy this situation, hybridization takes place and InTe forms a metallic band structure with a mixed ionic character at and near the Fermi level \EF, cf.\ Fig.~\ref{fig1}(c). Due to the significant In\,--\,Te hybridization, many bands exhibit sizable bonding-antibonding dispersions. Accordingly, this creates several energy valleys, appearing as separated electron and hole pockets in the BZ of InTe as sketched in Fig.~\ref{fig1}(d). The only electron pocket is centered at the $\Gamma$ point, whereas the major hole pockets span the BZ around its L points.

\begin{figure}[t]
\centering
\includegraphics[width=1\linewidth]{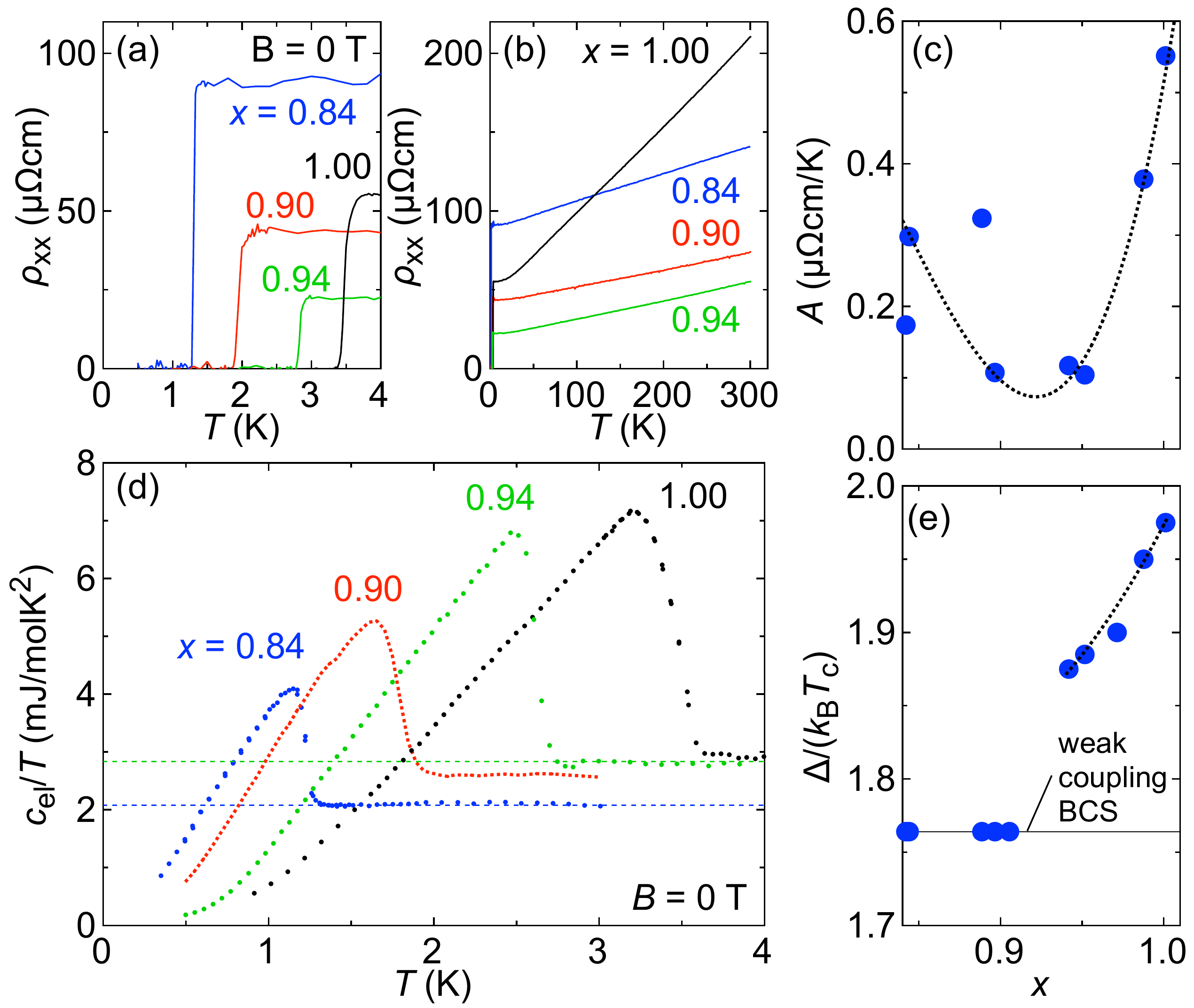}
\caption{(a) Resistivity \rxx\ of selected \IT\ samples with $0.84 \leq x \leq 1$ for $T\leq 4$~K, (b) \rxx\ up to room temperature exhibiting a characteristic linear temperature dependence for $\sim 40$~K $< T \leq 300$~K. (c) Slope $A$ of the $T$-linear part of $\rxx(T)$ as a function of $x$. (d) Electronic specific-heat data displayed as $\cel/T$ vs $T$. The dashed horizontal lines indicate the normal-state electronic specific-heat coefficient \gn\ for $x=0.84$ (blue) and 0.94 (green). Fits to these data in a BCS framework are shown in Fig.~S6 in \cite{Suppl}. From these, the normalized superconducting gap $\alpha=\Delta/\kB\Tc$ was extracted, and plotted against $x$ in (e), see text. Therein the solid horizontal line indicates the weak-coupling BCS limit $\alpha = 1.764$. Dotted lines in (c) and (e) are guides to the eyes.}
\label{fig2}
\end{figure}
Figure~\ref{fig2} summarizes resistivity and specific-heat data along with fitting results of these quantities on selected samples \IT\ with $0.84 \leq x \leq 1$. Resistivity data are presented in Figs.~\ref{fig2}(a) and (b) for $T\leq 4$~K and up to room temperature, respectively. We find sharp drops to zero resistivity for all samples examined. In the normal state, all samples exhibit a linear temperature dependence over a wide temperature range $\sim 40~{\rm K}<T \leq 300$~K. The $T$-linear slope $A$ estimated from fits $\rxx(T)\propto AT$ to the data in this temperature range is plotted as a function of $x$ in Fig.~\ref{fig2}(c). Although there is some scatter among the data points in the low-$x$ region, $A$ exhibits a steep increase for $x\gtrsim0.92$, clearly indicating an enhancement of the electron-phonon scattering toward stoichiometric InTe.

Figure~\ref{fig2}(d) presents zero-field electronic specific-heat data displayed as $\cel/T$ vs $T$ for $T \leq 4$~K for selected samples. All samples exhibit a clear and sharp jump-like anomaly at \Tc, indicating bulk superconductivity in \IT. The lower and upper dotted horizontal lines represent the normal-state electronic specific-heat coefficient \gn\ for $x = 0.84$ (blue data) and $x = 0.94$ (green), respectively. Apparently there is a strong enhancement of \gn\ when increasing the In content from $x=0.84$, but this enhancement saturates for $x >0.9$. For larger $x$ there is no appreciable increase any more. The electronic specific heat is further analyzed in a BCS framework employing the $\alpha$ model \cite{muehlschlegel59a,padamsee73a}, see Section~S4 in \cite{Suppl}. Herein, $\alpha= \Delta/\kB\Tc$ is a measure of the superconducting coupling strength with the superconducting energy gap $\Delta$ (at 0~K), cf.\ Section~S4 in \cite{Suppl} for a detailed description of this approach. Figure~\ref{fig2}(e) summarizes $\alpha(x)$.  In accord with the observed changes in the slope $A$ of the linear resistivity and the flattening of \gn, $\alpha$ starts to increase beyond the weak-coupling BCS limit for $x \gtrsim 0.92$, reaching almost $\alpha = 2$ when $x$ approaches $1$. The most striking experimental result here is the clear change in the superconducting response of the system across $x \sim 0.92$: $A$, $\alpha$, and \Tc\ increase toward $x=1$, while \gn\ saturates, suggesting a change in the dominant ingredients governing the superconductivity in \IT. 

\section{Theory}
\label{theory}

\begin{figure}[t]
\centering
\includegraphics[width=1\linewidth]{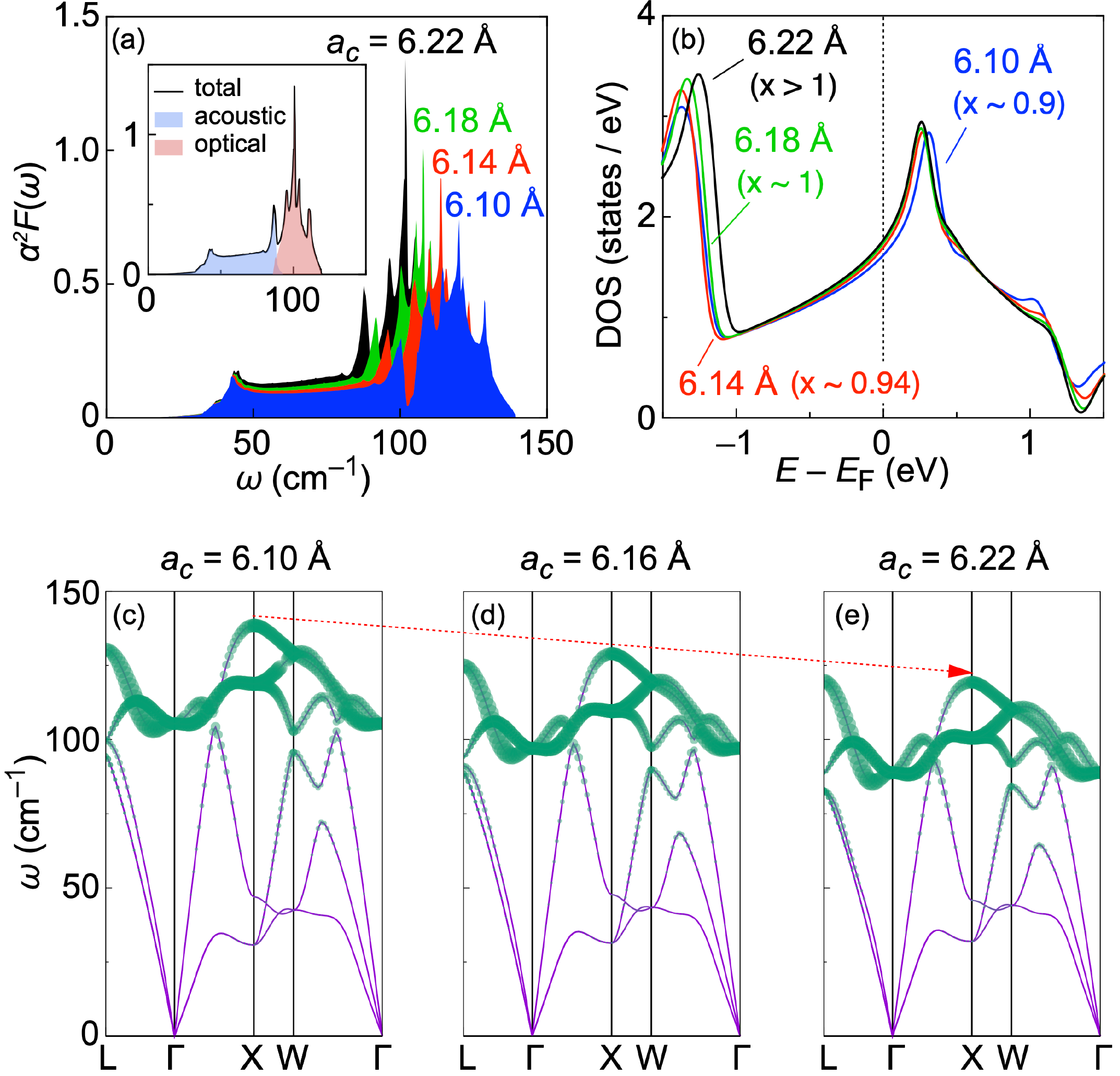}
\caption{(a) Phonon frequency dependence of the Eliashberg function $\alpha^2F(\omega)$ for selected lattice constants $a_c = 6.10$~\AA\ (blue), 6.14~\AA\ (red), 6.18~\AA\ (green), and 6.22~\AA\ (black). The inset shows exemplarily for $a_c = 6.22$~\AA\ how the spectral weight of $\alpha^2F$ distributes to acoustic (blue shading) and optical phonons (red). (b) Electronic density of states (DOS) given in states / eV and spin direction calculated for the same lattice constants of InTe as in (a). Corresponding $x$ values are indicated in brackets. The dotted vertical line indicates the Fermi energy \EF\ for InTe. Energy dispersion of phonon frequencies $\omega$ along high-symmetry lines in $k$ space for selected lattice constants (c) $a_c= 6.10$~\AA, (d) 6.16~\AA, and (e) 6.22~\AA. The strength of electron-phonon coupling \lamep\ associated with each $k$ point is shown by green dots. The larger are the dots the stronger is \lamep. The drastic change of the phonon frequencies with $a_c$ is highlighted by the red arrow, see text.}
\label{fig3}
\end{figure}

To shed light on this issue, we have developed a mean-field model which successfully reproduces our observations for $x \gtrsim 0.92$. The results are summarized in Fig.~\ref{fig3}. Starting with Fig.~\ref{fig3}(a), we have calculated the Eliashberg function $\alpha^2 F(\omega)$ as a function of the phonon frequency $\omega$ for InTe with cubic lattice constants $a_c=6.10$~\AA\ (blue), 6.14~\AA\ (red), 6.18~\AA\ (green), and hypothetical 6.22~\AA\ (black), corresponding to $x\sim 0.9$, 0.94, 1, and $x>1$, respectively. The spectral weight of $\alpha^2 F$ increases with $a_c$, already suggesting a possible increase in \lamep\ upon expanding the crystal lattice of \IT, which will become clearer in the next Section. As highlighted in the inset, the optical phonons are the main source of this enhancement. Figure~\ref{fig3}(b) shows the calculated electronic DOS data for the same lattice constants. 

Figures~\ref{fig3}(c)\,--\,(e) summarize the calculated energy dispersions of phonon frequencies in InTe for $a_c=6.10$~\AA, 6.16~\AA, and 6.22~\AA, respectively. As indicated by the red arrow, the optical phonon frequencies, which are the dominant source of \lamep, are strongly softened upon increasing $a_c$. This points towards a scenario of the superconductivity where the lattice expansion in \IT\ may play a significant role.

\section{Discussion}
\label{discussion}
\begin{figure}[t]
\centering
\includegraphics[width=1\linewidth]{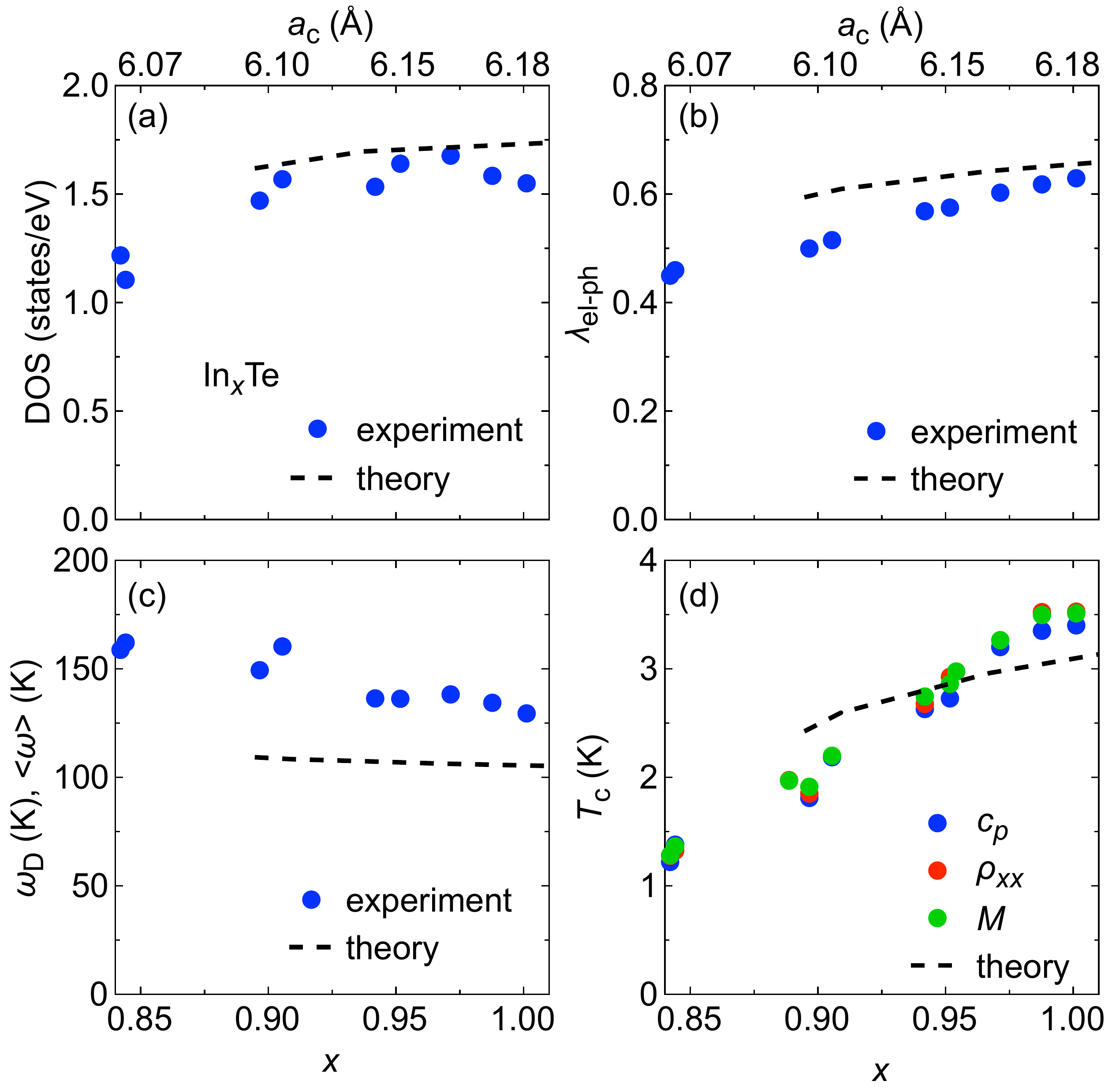}
\caption{(a) Density of states (DOS) (experimental data are corrected for electron-phonon interaction, see text), (b) electron-phonon coupling parameter \lamep, (c) Debye frequency \omD\ (from \sh) and averaged phonon frequency $\textless\omega\textgreater$ (from theory), and (d) superconducting \Tc\ as functions of $x$. For comparison, the corresponding approximate lattice constants $a_c$ are shown on the top axes of panel (a) and (b). Blue data points in all panels are estimated from specific-heat, red and green data in (d) from  resistivity (\rxx) and magnetization ($M$) data, respectively. Black dashed lines in all panels indicate the results of our theoretical model calculations. The apparent enhancement of DOS by $\sim 0.5$~states/eV below $x\sim 0.92$ is discernible in (a).}
\label{fig4}
\end{figure}

Figure~\ref{fig4} compares experimental and theoretical data of \IT. Blue data points in all panels are estimated from specific-heat data. Red and green data points in Fig.~\ref{fig4}(d) refer to \Tc\ values deduced from resistivity and magnetization data, respectively. The procedure how experimental DOS, \lamep, and \omD\ data were deduced from specific-heat measurements is described in Section S4 in \cite{Suppl}. The theoretical values are calculated as a function of the lattice constant $a_c$ and are plotted in Fig.~\ref{fig4} as black dashed lines against the corresponding In concentration $x$. 

Density-of-states data are shown in Fig.~\ref{fig4}(a) as a function of $x$. Since theoretical calculations yield the bare DOS without electron-phonon interaction, therein the experimental DOS results are divided by $1+\lamep$ to allow a quantitative comparison. In accordance with the electronic specific-heat data shown in Fig.~\ref{fig2}(d), there is an enhancement of DOS upon increasing $x$ from 0.84 to 0.92, which amounts to almost $\sim 0.5$~states/eV. However, above $x\sim 0.92$, the enhancement becomes gradual and DOS seems to saturate. The latter is very well reproduced by our theoretical modeling: the black dashed line is DOS from Fig.~\ref{fig3}(b) replotted as a function of $x$, matching well with the experimental data. Figure~\ref{fig4}(b) shows the corresponding electron-phonon coupling strength \lamep, which increases from 0.45 to 0.63 when varying the In content from $x=0.84$ to 1.0. The black dotted line therein represents integrated Eliashberg functions $\lamep = 2\int d\omega\, \alpha^2F(\omega)/\omega$ \cite{allen75a} shown in Fig.~\ref{fig3}(a) which again agrees well with experimental data.

The $x$ dependences of the experimental Debye frequency \omD\ and its theoretical counterpart $\textless\omega\textgreater = 2/\lamep \int d\omega \alpha^2 F(\omega)$ \cite{allen75a} are given in Fig.~\ref{fig4}(c). Experimentally we observe a decrease of \omD\ with $x$, indicating a softening of the lattice vibrations as suggested by Figs.~\ref{fig3}(c)\,--\,(e). The negative slope of $\omD(x)$ is reproduced in our model calculations.

Theoretical and experimental \Tc\ values are summarized in Fig.~\ref{fig4}(d). For $x=0.84$, the smallest $\Tc\sim 1.3$~K in this study is found \cite{commentTc}. Upon  increasing $x$, \Tc\ enhances linearly in accord with Refs.~\cite{geller64a,geller64b}. Close to stoichiometric InTe, the slope decreases and \Tc\ seems to saturate. Again, our theoretical model catches the \Tc\ values for $x>0.92$ well. 

Finally, we will propose a possible scenario of this change in the superconductivity in \IT. Below $x \sim 0.92$, an interpretation of our data in terms of a mainly DOS-driven enhancement of \lamep\ and \Tc\ seems obvious, as suggested by the concomitant increase of DOS with $x$, when comparing Figs.~\ref{fig4}(a), (b), and (d). However, above $x\sim 0.92$, DOS flattens while \lamep\ and \Tc\ keep increasing. The continuous increase of \lamep\ is ascribed to the softening of phonons in the higher $x$ region.

A phenomenological explanation can be given when considering the observed expansion of the cubic lattice with $x$ as shown in Fig.~\ref{fig1}(b). The lattice expansion is reasonable because more and more voids in the InTe matrix get filled with $x$. The shorter lattice constants at low $x$ imply a more rigid lattice than for larger $x$. As a consequence, the electron-phonon interaction is relatively weak at small $x$, and therefore, \Tc\ is small. Upon filling voids in the In sublattice, the crystal lattice expands and atoms are shifted more apart from each other. This allows for an easier vibration of the constituent atoms, leading to softer phonon modes as seen in Figs.~\ref{fig3}(c) to (e). This can be regarded as a ``negative pressure'' effect (lattice expansion) in analogy to the physical pressure effect (lattice contraction). Note that changes in \lamep\ and \Tc\ are induced mainly by DOS changes in conventional cases while these are governed by phonon softening in the present case. A similar phonon-softening induced enhancement of superconductivity is discussed in literature \cite{kudo12a}.

\begin{figure}[t]
\centering
\includegraphics[width=0.9\linewidth]{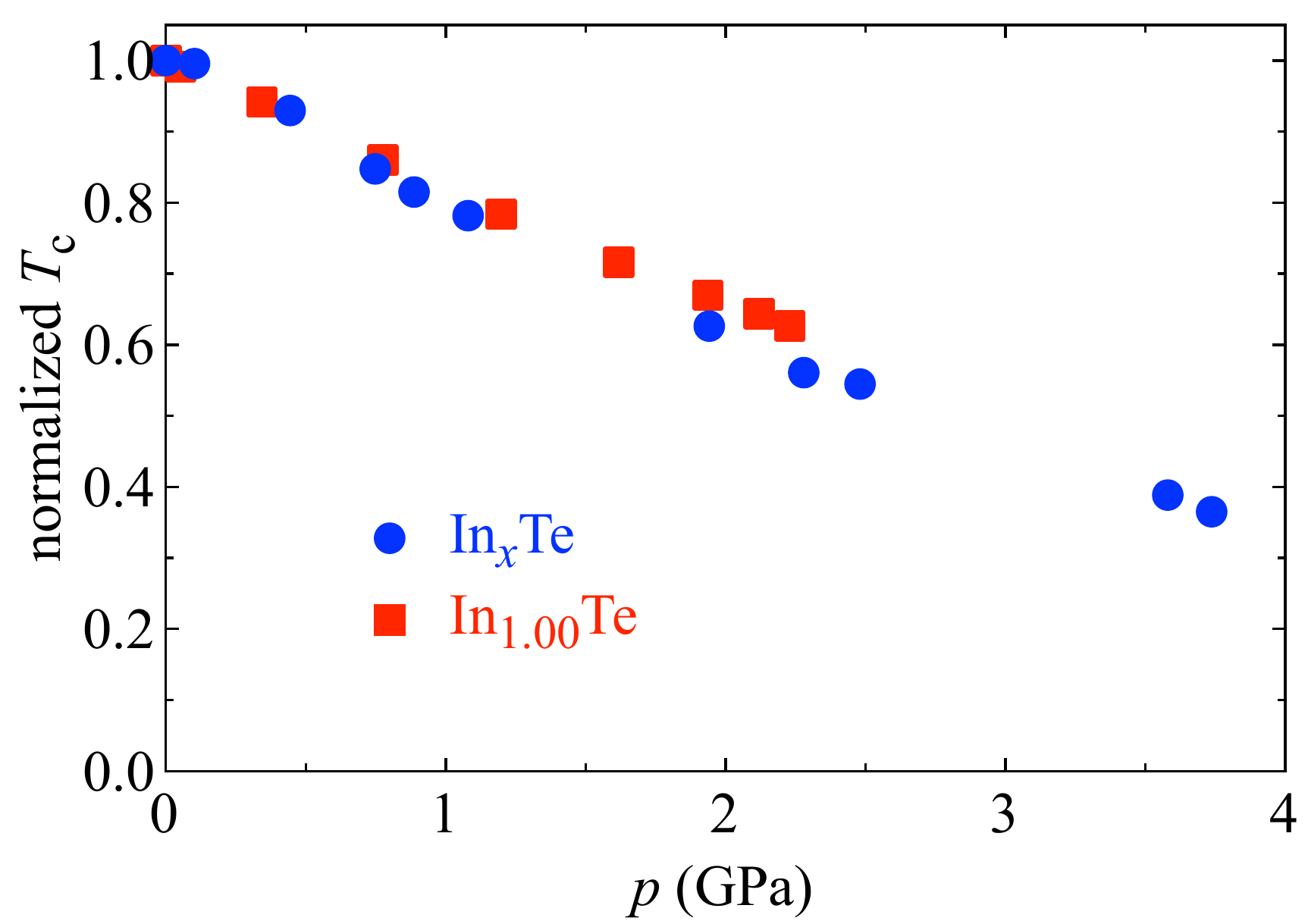}
\caption{Comparison of lattice-constant-dependent \Tc\ in \IT\ (blue circles) and physical pressure effect on a sample with $x=1.00$ (red squares). For In-deficient \IT, the change in the lattice constant is converted to physical pressure by using the bulk modulus for cubic InTe, i.e., $x$ decreases as the converted pressure value is increased, see text. The \Tc\ values are normalized with respect to the largest $\Tc\sim 3.5$~K observed for $x=1.00$ in this work.}
\label{fig5}
\end{figure}

To further test this hypothesis, we converted the ``negative pressure'' effect into physical pressure, which is shown with blue circle symbols in Fig.~\ref{fig5}. The \Tc\ values are normalized with respect to the largest $\Tc=3.51$~K (for $x=1.00$; all \Tc\ values are from magnetization measurements) and plotted against physical pressure $p$. The latter values are calculated from the change of each sample's unit-cell volume $V$ again with respect to the sample with $x=1.00$ by using the known bulk modulus \cite{chattopadhyay85a}, cf.\ Section~S5 in \cite{Suppl} for a detailed description of the conversion process. For comparison, we also determined the phyical pressure dependence $\Tc(p)$ from resistivity measurements on a sample with $x=1.00$. These data are shown with red square symbols in Fig.~\ref{fig5} and are normalized to the ambient-pressure zero-resistance $\Tc=3.53$~K of this sample, cf.\ Section~S6 in \cite{Suppl}. The agreement between converted $\Tc(x)$ and measured $\Tc(p)$ is very good and indeed pointing towards a scenario where the lattice expansion plays a significant role.

The total change of \Tc\ in \IT\ for $0.84\leq x \leq 1.00$ corresponds to an applied physical pressure of $p\sim 3.75$~GPa. To further contextualize this value, a comparison between the case of \IT\ and several superconducting elements is shown in Section~S5 in \cite{Suppl}. It turns out that \IT\ exhibits one of the strongest pressure effects observed in such chemically simple BCS superconductors, emphasizing that the soft lattice in \IT\ is indeed in favor of its superconductivity. This may be related to the valence-fluctuation feature of In in the present compound which could be an interesting starting point for future studies.

\section{Summary}
\label{summary}
To summarize, we demonstrate that controlling and fine-tuning superconductivity can be achieved by simply changing the In concentration $x$ in \IT. For $x < 0.92$, the enhancement of the superconductivity can be straight-forwardly understood as a consequence of an enhanced density of states at the Fermi level. Upon further increasing $x$, this enhancement fades out and the observed continued increase of the superconducting \Tc\ cannot be attributed exclusively to a density-of-states effect any more. Therefore, we modeled this In concentration range theoretically and reproduced well the experimental data of density of states, electron-phonon coupling constant, and superconducting transition temperature for $0.92 < x \leq 1$ by solely changing the cubic lattice constant. Together with the outcome of our physical pressure experiment, our results suggest that the enhanced superconductivity in this system is closely related to the soft lattice of this simple chalcogenide.

\section*{Acknowledgement}
This work was partly supported by Grants-In-Aid for Scientific Research (S) from the Japan Society for the Promotion of Science (JSPS, No.\ 24224009), JST (No.\ JP16H00924), and PRESTO (JPMJPR15N5) and Grants-In-Aid for Scientific Research (B) (JSPS, No.\ 17H02770). We thank the RIKEN Materials Characterization Team for compositional analyses. M.K. thanks D.~Hashizume, T.~Kikitsu, D.~Inoue, T.~Nakajima, and D.~Maryenko for fruitful discussions.
M.K. and M.S.B. contributed equally to this work.


%

 

\begin{thebibliography}{29}
\bibitem [{com()}]{commentTc}%
  \BibitemOpen
  \href@noop {} {}\bibinfo {note} {{As mentioned in Section~\ref{methods}, we
  also synthesized one sample with a smaller $x = 0.79$ which contains
  additional minority phases. The \Tc\ of its superconducting main phase
  fraction is comparable to the \Tc\ of the sample with $x=0.84$ suggesting
  that \Tc\ does not further decrease due to the apparent border of stability
  of cubic \IT\ in this $x$ range.}}\BibitemShut {Stop}%
\end{thebibliography}

\newpage

\end{document}


\title{Supporting Materials for \\ ``Enhancement of superconductivity and its relation to lattice expansion in InTe''}

\author{M.~Kriener}
\email[corresponding author: ]{markus.kriener@riken.jp}
\affiliation{RIKEN Center for Emergent Matter Science (CEMS), Wako 351-0198, Japan}
\author{M.~S.~Bahramy}
\email[corresponding author: ]{m.saeed.bahramy@manchester.ac.uk}
\affiliation{Department of Physics and Astronomy, The University of Manchester, Oxford Road, Manchester M13 9PL, United Kingdom}
\author{Y.~Tokura}
\affiliation{RIKEN Center for Emergent Matter Science (CEMS), Wako 351-0198, Japan}
\affiliation{Department of Applied Physics and Quantum-Phase Electronics Center (QPEC), University of Tokyo, Tokyo 113-8656, Japan}
\affiliation{Tokyo College, University of Tokyo, Tokyo 113-8656, Japan}
\author{Y.~Taguchi}
\affiliation{RIKEN Center for Emergent Matter Science (CEMS), Wako 351-0198, Japan}

\date{\today}

\maketitle

This Supplement provides additional data on \IT.

\begin{itemize}
\item Section~\ref{xrd}: Crystal structure at room temperature
\item Section~\ref{icp}: Results of ICP analyses
\item Section~\ref{Mag}: Magnetization
\item Section~\ref{cpInTe}: Specific heat analyses
\item Section~\ref{negP}: ``Negative pressure'' effect in \IT\ compared to simple BCS superconductors
\item Section~\ref{Tcp}: Physical pressure effect measured on a sample In$_{1.00}$Te
\end{itemize}

\clearpage

\subsection{Crystal structure at room temperature}
\label{xrd}
\begin{figure}[h!]
\centering
\includegraphics[width=0.85\linewidth,clip]{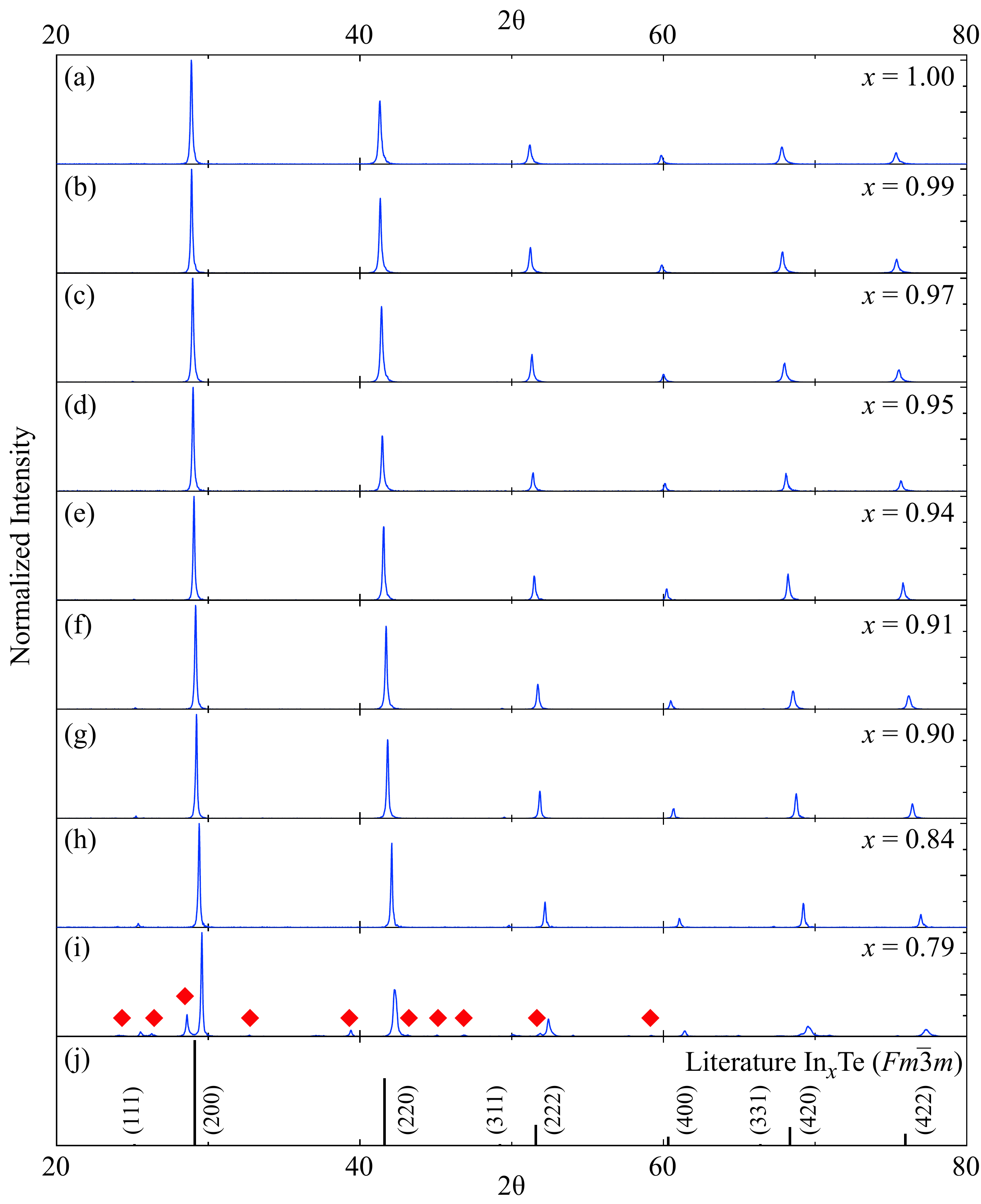}
\caption{Normalized XRD data of \IT\ batches with $0.79\leq x \leq 1.00$. The lowermost panel shows the expected peak positions and their intensities for cubic InTe. Labels indicate the respective $hkl$ indices. Only the batch with the lowest In concentration $x = 0.79$ examined contains clear impurity phase fractions whose respective peaks are indicated with red diamond symbols.}
\label{figS1}
\end{figure}

\subsubsection*{\textbf{XRD patterns}}
Figure~\ref{figS1} summarizes XRD data for various batches of \IT\ ($0.79\leq x \leq 1.00$) normalized with respect to the 200 main peak of the cubic InTe structure (space group 225, $Fm\bar{3}m$). Panel (j) depicts the expected peak positions for the lattice constant $a_c = 6.133$~\AA\ taken from literature \cite{faita11a}. This $a_c$ corresponds to $x \approx 0.93$. The peaks are also labeled therein. All peaks in the experimental diffraction patterns are sharp (cf.\ the discussion below). There are no detectable impurity phases within the resolution of the diffractometer for $x > 0.85$. The batch with the lowest In concentration $x = 0.79$ contains about 20\% impurity phases In$_y$Te$_z$ as estimated from the intensity comparison of the largest impurity peak on the left side of the 200 InTe main peak. Additional peaks due to impurity phases are marked with diamond symbols, cf.\ Fig.~\ref{figS1}(i). As for the batch with $x = 0.84$ [Fig.~\ref{figS1}(h)], only very tiny traces of impurity contributions are recognizable. This suggests that the cubic InTe phase becomes unstable around $x \sim 0.8$. Hence, in the main text, the discussion is focussed on the In concentration range $0.84 \leq x \leq 1.00$.

As can be also seen in Fig.~\ref{figS1}, there are clear peak shifts toward larger $2\theta$ values when going from panel (a) to (h), indicating a monotonic increase of the cubic lattice constant $a_c$ as a function of increasing $x$, see Fig.~1(b) in the main text. 

\subsubsection*{\textbf{XRD peak width}}
\begin{figure}[h]
\centering
\includegraphics[width=12cm,clip]{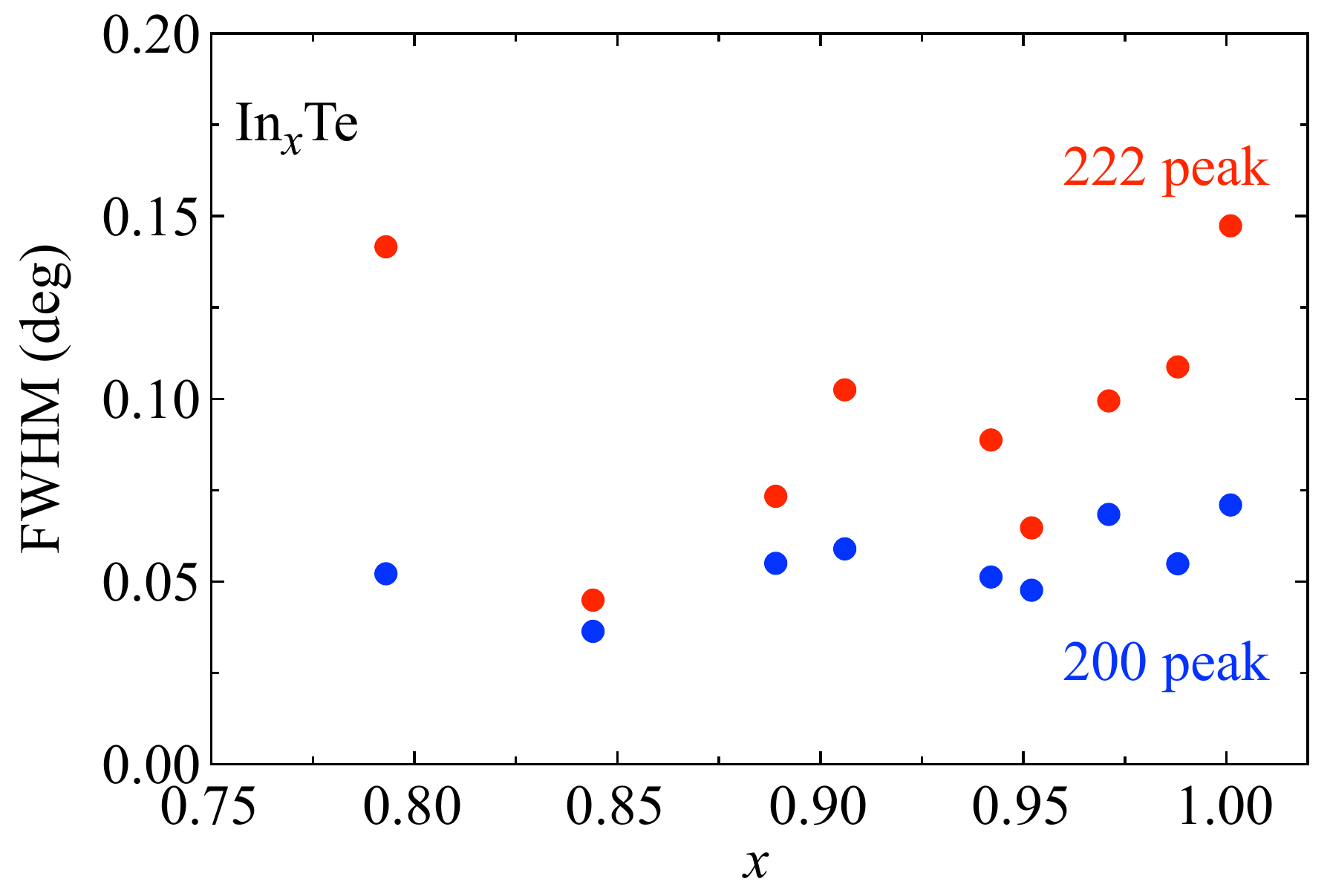}
\caption{Full-width-half-maximum (FWHM) of the cubic 200 (blue data) and 222 reflections (red) as a function of the In concentration $x$. The peaks are fairly sharp indicating a good sample quality but the peak width gradually increases with $x$.}
\label{figS2}
\end{figure}
Figure~\ref{figS2} shows exemplarily the $x$ dependence of the full-width-half-maximum (FWHM) of the cubic 200 (main peak) and 222 reflections. Apparently they are fairly sharp, implying high sample quality. However, there is a slight broadening of the peaks visible below $x = 0.84$ and towards stoichiometric InTe. In the case of the sample with the smallest $x=0.79$ this is probably a consequence of the impurity phase formation. As for the slight broadening of the peaks toward $x=1$, this may indicate that the grain size decreases toward $x=1$ under the present synthesis condition.

\newpage
\subsubsection*{\textbf{Possible formation of Te antisite defects}}
Another point to be checked is whether some Te ions move into In voids, forming antisite disorder which is conceivable especially at larger $x$. To address this issue we analyzed the peak intensities of the 200 main peak ($2\theta \sim 29^{\circ}$) and the 111 peak ($2\theta \sim 25^{\circ}$) as follows (cf., e.g., \cite{ashcroft76}). The peak intensity $I$ is proportional to the square of the structure factor $F_{hkl}$ ($hkl$ denote the Miller indices):
\begin{equation}
\label{inten}
I \propto |F_{hkl}|^2.
\end{equation} 
In turn $F_{hkl}$ of an fcc lattice (as it is the case in InTe) is given by
\begin{equation}
\label{strfacfcc}
 F_{hkl}^{\rm fcc}=\sum_{j} \exp\left[2\pi i\,\vec{d_j}\cdot \vec{b}\right],
\end{equation} 
with summation over all atoms in the unit cell. Here $\vec{d}_j=(a\;b\;c)$ and $\vec{b}=(h\;k\;l)$ are the position and the reciprocal lattice vectors of the $j^{\rm th}$ atom, respectively. In an fcc lattice, the conventional unit cell is built up by four atoms at $\vec{d_1} = (0\;0\;0)$, $\vec{d_2} = (0.5\;0.5\;0)$, $\vec{d_3} = (0\;0.5\;0.5)$, and $\vec{d_4} = (0.5\;0\;0.5)$ and $F_{hkl}^{\rm fcc}$ evaluates to 4 if $h$, $k$, and $l$ are all even or all odd and 0 otherwise.

InTe consists of two fcc lattices shifted by $\Delta \vec{d} = (0.5\;0.5\;0.5)$, i.e., $\vec{d_j}^{\rm Te}=\vec{d_j}^{\rm In} +\Delta \vec{d}$. Hence, the total structure factor of InTe amounts to
\begin{eqnarray}
\label{strfac}
 F_{hkl}&=&g_{\rm In} f_{\rm In} \sum_{j\,\in\, \rm In}\exp\left[2\pi i\,\vec{d_j}^{\rm In}\cdot \vec{b}\right] + g_{\rm Te} f_{\rm Te} \sum_{j\,\in\, \rm Te}\exp\left[2\pi i\,\vec{d_j}^{\rm Te}\cdot \vec{b}\right]\\
 & = & g_{\rm In} f_{\rm In} F_{hkl}^{\rm fcc} + g_{\rm Te} f_{\rm Te} F_{hkl}^{\rm fcc} \exp\left[2\pi i\,\Delta\vec{d}\cdot \vec{b}\right]\\
 & = & \left( g_{\rm In} f_{\rm In} + g_{\rm Te} f_{\rm Te} \exp\left[2\pi i\,\frac{h+k+l}{2}\right] \right)F_{hkl}^{\rm fcc}, 
\end{eqnarray} 
with the site occupancies $g_k$ and atomic form factors $f_k$ of $k =$~In and Te. The exponential function evaluates to $+1$ when $h+k+l$ is even and $-1$ when it is odd, hence, $F_{hkl}\propto (g_{\rm In} f_{\rm In} \pm g_{\rm Te} f_{\rm Te})\cdot F_{hkl}^{\rm fcc}$. For the 111 and 200 reflections, the atomic form factors of In and Te are similar (within 5\%). Therefore, for stoichiometric InTe $g_{\rm In} = g_{\rm Te} = 1$, the intensity of the ``all odd" reflection 111 should be much smaller than 1 while the intensity of the ``all even'' reflection 200 should be much larger than 1 in agreement with the experimental observation that the 200 peak is strong and its neighboring 111 peak is weak. When changing $x$ and, hence, $g_{\rm In} < g_{\rm Te} = 1$, the ``all odd'' peak intensity should increase. This is again in qualitative agreement with Fig.~\ref{figS1}: Upon reducing $x$, the 111 peak intensity becomes larger. To quantify this, we calculated the intensity ratio of these two peaks for our \IT\ samples. If Te remains on its regular lattice sites in In-deficient \IT, i.e., $g_{\rm Te} = 1$ and assuming $f_{\rm In} \cong f_{\rm Te}$ then the intensity ratio is approximately (with $F_{111}^{\rm fcc} = F_{200}^{\rm fcc} = 4$) 
\begin{equation}
\label{strfacInTe}
\frac{I_{111}}{I_{200}} = \frac{|g_{\rm In} f_{\rm In} - g_{\rm Te} f_{\rm Te}|^2}{|g_{\rm In} f_{\rm In} + g_{\rm Te} f_{\rm Te}|^2}\cdot \frac{|F_{111}^{\rm fcc}|^2}{|F_{200}^{\rm fcc}|^2} 
 =\frac{(g_{\rm In}-1)^2}{(g_{\rm In}+1)^2}\propto (g_{\rm In}-1)^2
\end{equation}. 

We note, that in reality the intensity ratio will not become exactly 0 even for $g_{\rm In} = x = 1$ because the atomic form factors are similar but not identical. Hence, the 111 reflection exists in stoichiometric InTe, although possibly not visible depending on the x-ray intensity of the diffractometer used. 

\begin{figure}[h]
\centering
\includegraphics[width=12cm,clip]{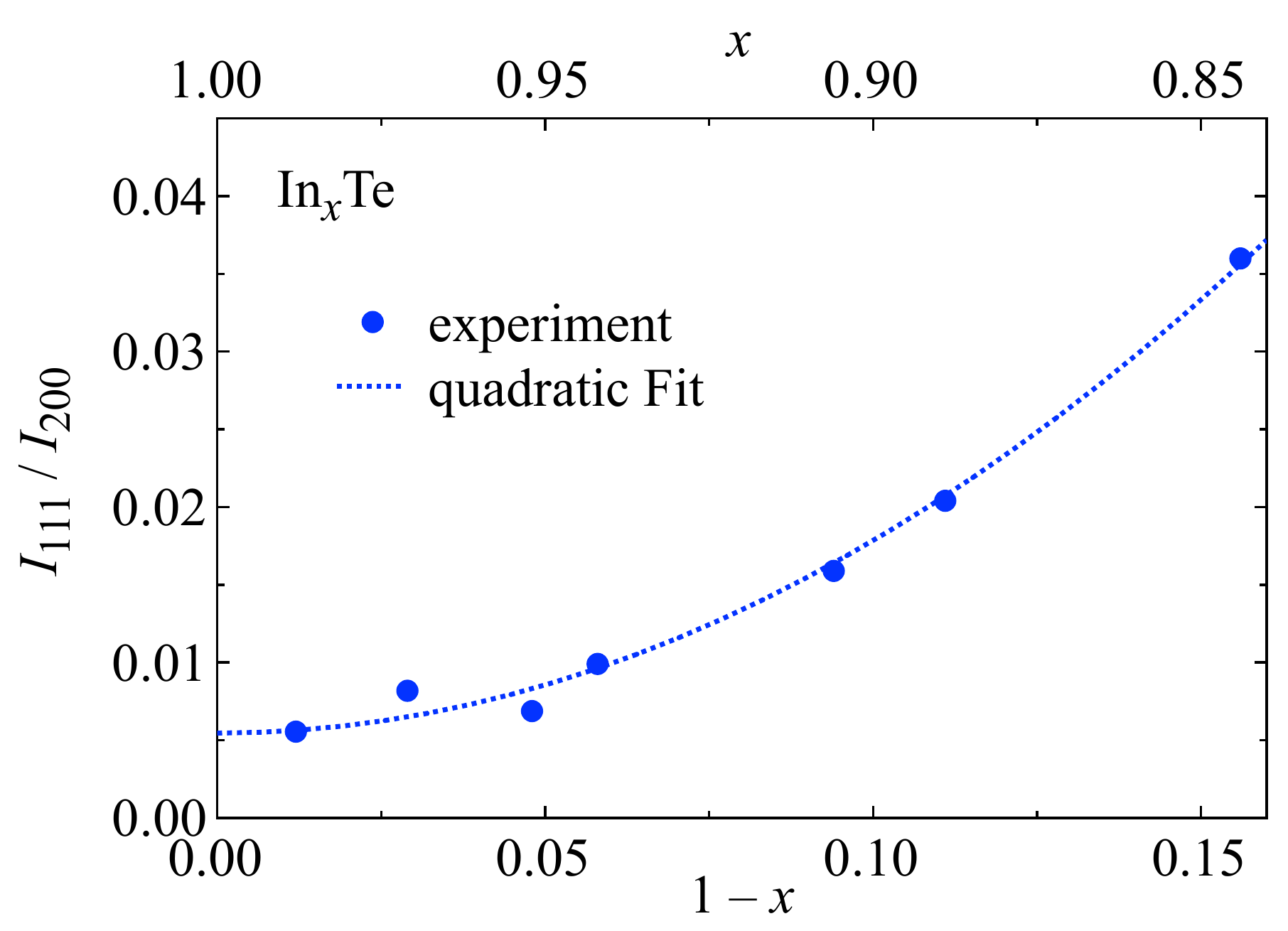}
\caption{Ratio of the intensities of the 111 and 200 reflections as a function of $1-x$. If Te would enter In voids at larger $1-x$, this ratio should fall below the expected quadratic behavior, see text for details.}
\label{figS3}
\end{figure}

The quadratic dependence in Eq.~\ref{strfacInTe} holds in our case as shown in Fig.~\ref{figS3}. This clearly demonstrates that Te remains on its regular lattice sites, and does not form antisite disorder. We also note, that strictly speaking one has to take into account the integrated intensities here. This is neglected because both peaks are close to each other and at small $2\theta$ values where the instrument broadening effect does not cause big differences.

\clearpage

\subsection{Results of ICP analyses}
\label{icp}
\begin{figure}[h]
\centering
\includegraphics[width=10cm,clip]{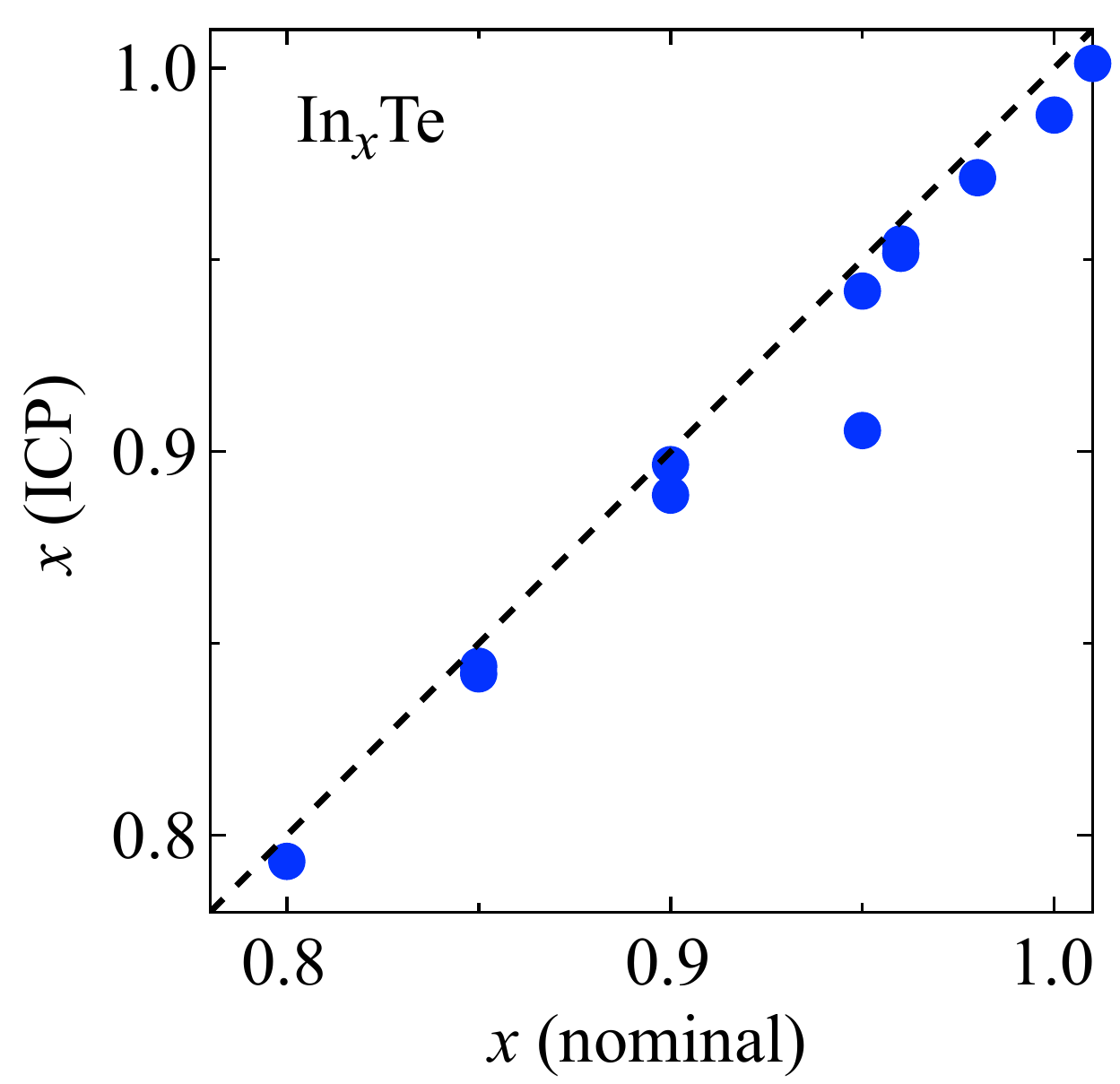}
\caption{Result of ICP analyses of \IT.}
\label{figS4}
\end{figure}

Results of ICP analyses on \IT\ are shown in Fig.~\ref{figS4}. Except for one case, these are close to the nominal values indicated by the dashed line. These ICP results are used as $x$ when referring to a sample.

\newpage

\subsection{Magnetization}
\label{Mag}
\begin{figure}[h]
\centering
\includegraphics[width=12cm,clip]{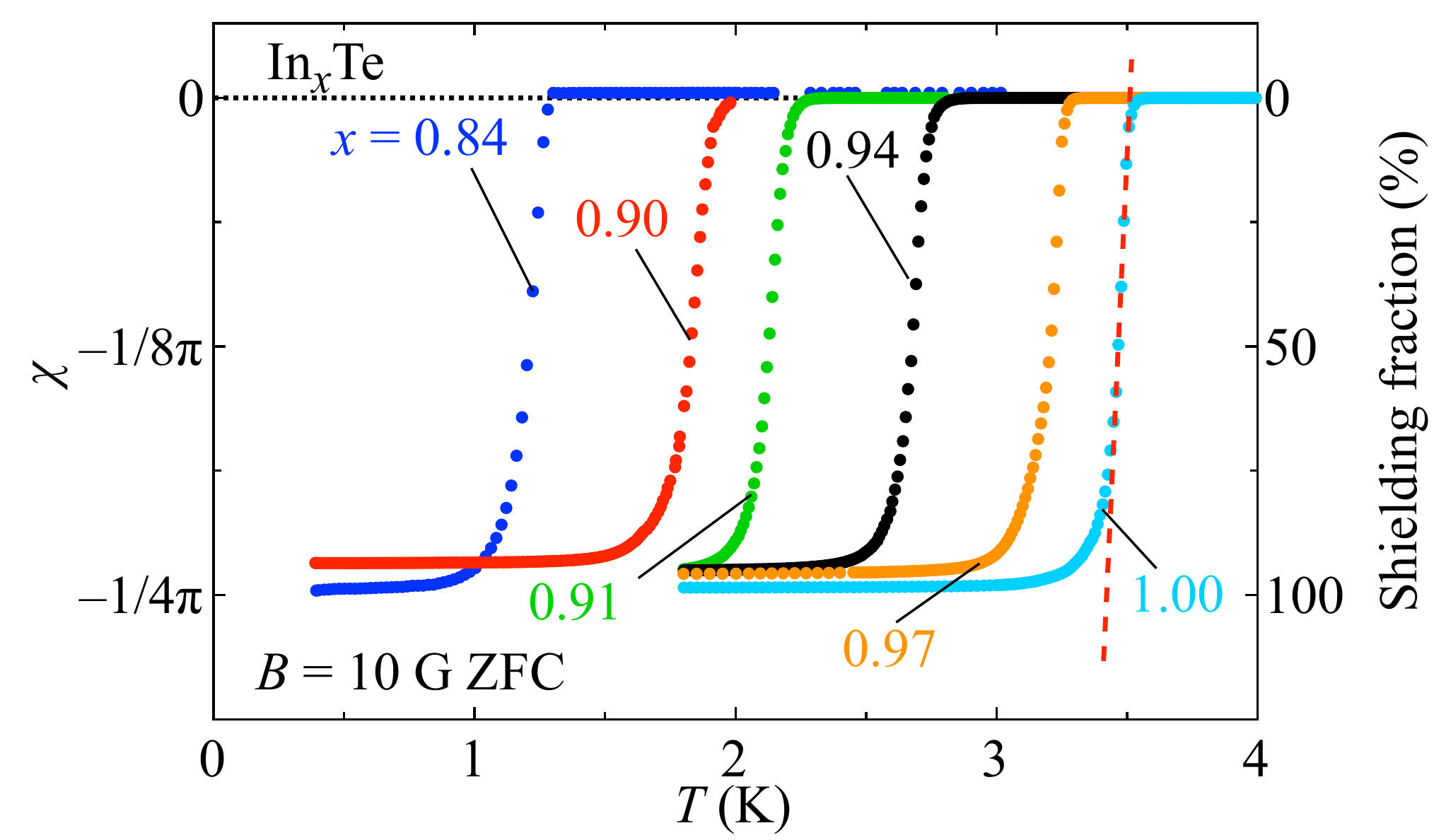}
\caption{Susceptibility and corresponding superconducting shielding fraction of \IT\ corrected for the demagnetization effect, see text. Superconducting \Tc\ values are defined as intersection of a linear fit to data in the transition range of a sample and the normal-state base line, as exemplarily shown by the dashed red fit and dotted black baseline for In$_{1.00}$Te.}
\label{figS5}
\end{figure}

Figure~\ref{figS5} summarizes the superconducting shielding signal for various samples of \IT\ as estimated from temperature-dependent magnetization data. The data has been corrected for the demagnetization effect by employing Brandt's formula for samples in slab-like geometry \cite{brandt99a}.  The slight deviations from 100\% are due to several effects: The superconducting magnet of the magnetometer used exhibits a small but unknown remanent field, deviations from an exact slab geometry, and errors in the sample volume which was estimated from each sample's linear dimensions. The red dashed line is a linear fit to the data for $x = 1.00$ in its transition range. 

\clearpage

\subsection{Specific heat}
\label{cpInTe}

\subsubsection*{\textbf{Measurements}}
\begin{figure}[h]
\centering
\includegraphics[width=13.25cm,clip]{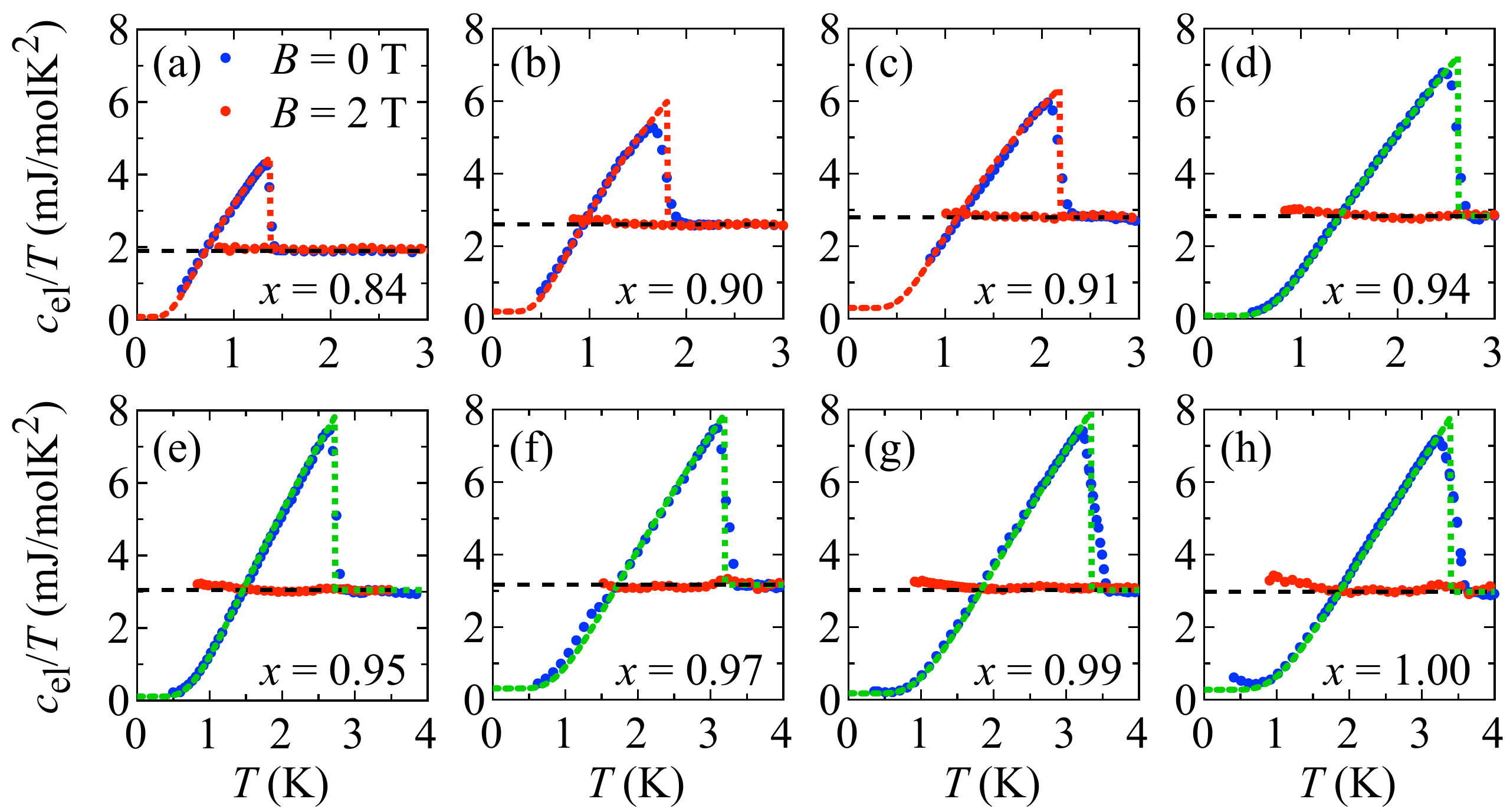}
\caption{Electronic specific-heat data for various samples \IT\ with $0.84 \leq x \leq 1$ displayed as $\cel/T$ vs $T$ in $B=0$~T (blue data) and 2~T (red). Dashed horizontal lines in each panel indicate the normal-state electronic specific-heat coefficient \gn, dotted curves are the electronic specific-heat in a weak-coupling [red; panels (a) -- (c)] or strong-coupling BCS framework [green; (d) -- (h)], cf.\ text.}
\label{figS6}
\end{figure}
Figure~\ref{figS6} summarizes electronic specific-heat data for various samples \IT\ with $0.84 \leq x \leq 1$ displayed as $\cel/T$ vs $T$. These were measured in $B=0$~T (blue data) and 2~T (red). Background (addenda) data were taken in each case before the sample was mounted and duly subtracted from the total signal sample + addenda. A magnetic field of 2~T is sufficient to suppress the superconductivity in \IT\ and, hence, the in-field data represents the normal-state specific heat \sh.

\subsubsection*{\textbf{Electronic specific heat}}
The following equation was fitted to the normal-state \sh\ (not shown):
\begin{equation}
\sh = \cel + \cph = \gn T + A_3 T^3 + A_5 T^5
\end{equation}
with $\cel=\gn T$ and $\cph = A_3 T^3 + A_5 T^5$ denoting the electronic and phononic contributions to \sh, respectively. We added a $T^5$ term to the conventional Debye-$T^3$ law to account for anharmonic effects which allowed better fit results. This approach yields the Sommerfeld parameter \gn\ and from the coefficient $A_3$ of the Debye term via 
\begin{equation}
A_3 = \frac{12}{5}\,\frac{\pi^4 N N_{\rm A} k_{\rm B}}{\Theta_{\rm D}^3}
\end{equation}
the Debye temperature $\Theta_{\rm D}$. Here, $N_{\rm A}$ and $k_{\rm B}$ denote the Avogadro and the Boltzmann constants, respectively, and $N=2$ is the number of atoms per formula unit. 
Subtracting \cph\ from \sh\ yields \cel. These data were further analyzed in the standard weak-coupling Bardeen-Cooper-Schrieffer (BCS) theory. All samples are bulk superconductors as indicated by superconducting volume fractions $> 90$\%. To account in our analysis for the small but existent residual normal-conducting contributions, we rewrote $\gn = \gs + \gres$ with \gs\ (\gres) representing the superconducting (normal conducting) density of states (DOS) of the samples below \Tc. The latter is the origin of a remaining offset in $\cel/T$ when approaching 0~K:
\begin{equation}\label{ctfit}
\cel/T=\gres + c_{\rm el}^{\rm BCS}(T)/T.
\end{equation} 
The tabulated BCS specific-heat data given in \cite{muehlschlegel59a} were modeled by fitting a polynomial $c_{\rm el}^{\rm BCS}$ to them. Subsequently, this polynomial was adjusted to match the experimental $\cel/T$ data for each sample by tuning \Tc\ and \gs\ according to Eq.~\ref{ctfit} yielding the dotted curves in Fig.~\ref{figS6}. Apparently the electronic specific heat of all samples is reproduced well within the BCS theory. The most interesting point here is that the BCS coupling strength in \IT\ changes with $x$ as indicated by the color of the dotted lines: red in Figs.~\ref{figS6}(a) -- (c) (weak coupling) and green in Figs.~\ref{figS6}(d) -- (h) (strong coupling), respectively. Red refers to standard weak-coupling BCS theory with $\Delta/(\kB\Tc) = 1.764$ ($\Delta$ denotes the superconducting energy gap at 0~K). To achieve satisfactory descriptions for $x\gtrsim 0.92$, this ratio, often denoted as $\alpha$ (``$\alpha$ model'', \cite{padamsee73a}), has to be systematically increased to almost $\alpha = 2$ for $x=1$. This is a measure of the coupling strength in the BCS theory and is shown in Fig.~2(e) of the main text as a function of $x$.

\subsubsection*{\textbf{Estimation of DOS, \boldmath{$\lamep$}, and \boldmath{$\omD$}}}
In Fig.~4 of the main text, experimental and theoretical data for DOS, \lamep, and \omD\ are compared with each other. The former were determined from specific-heat data:
According to the Drude-Sommerfeld theory, the normal-state electronic specific-heat coefficient \gn\ is related to the density of states (DOS) via (cf., e.g., \cite{ashcroft76})
\begin{equation}\label{DOSeq}
\textrm{DOS} = \frac{3\gn}{\pi^2k^2_{\rm B}}.
\end{equation}
To properly convert the units and taking into account that in the BCS theory of superconductivity, the DOS is counted for each spin direction separately (cf., e.g., \cite{tinkham96}), Eq.~\ref{DOSeq} becomes
\begin{equation}
\textrm{DOS} = \frac{1}{2}\cdot\frac{3\gn}{\pi^2k^2_{\rm B}}\cdot \frac{eV_0}{V_{\rm mol}}=\frac{1}{2}\cdot\frac{3\gn}{\pi^2k^2_{\rm B}}\cdot \frac{e t}{N_{\rm A}}
\end{equation}
with the elemental charge $e$, the unit-cell volume $V_0$, and the molar volume $V_{\rm mol}=V_0 N_{\rm A}/t$. In the latter, $t$ denotes the number of formula units per unit cell: $t=4$ for cubic InTe. 
However, the DOS obtained from this formula cannot be readily compared to theoretical data which yields the bare band-structure density of states without correlation effects. Hence, the experimental value has to be corrected for these effects. To do so, the experimental electron-phonon constant $\lambda$ is calculated via the McMillan formula \cite{mcmillan68a} by using the experimental values of \Tc\ and $\Theta_{\rm D}$:
\begin{equation}
\Tc=\frac{\Theta_{\rm D}}{1.45}\exp \left( -\frac{1.04(1+\lamep)}{\lamep - \mu^*(1 + 0.62 \lamep)} \right).
\end{equation}
As in our theoretical calculations, we used $\mu^* = 0.10$ for the repulsive Coulomb potential to determine \lamep. The final formula to estimate the band-structure DOS data from \sh\ becomes (cf., e.g., \cite{mcmillan68a})
\begin{equation}
\textrm{DOS} =\frac{6e\gn}{\pi^2k^2_{\rm B}N_{\rm A}(1+\lamep)}
\end{equation}
in units of states / eV and for one spin direction.
To obtain the experimental counterpart \omD\ to the average phonon frequency $\textless\omega\textgreater$, we used the Allen-Dynes formula \cite{allen75a}
\begin{equation}
\Tc=\frac{\omD}{1.20}\exp \left( -\frac{1.04(1+\lamep)}{\lamep - \mu^*(1 + 0.62 \lamep)} \right).
\end{equation}
with the experimental results for \Tc\ and \lamep.

\newpage
\subsection{``Negative pressure'' effect in In\boldmath{$_{x}$}Te compared to simple BCS superconductors}
\label{negP}

\begin{figure}[h]
\centering
\includegraphics[width=10cm,clip]{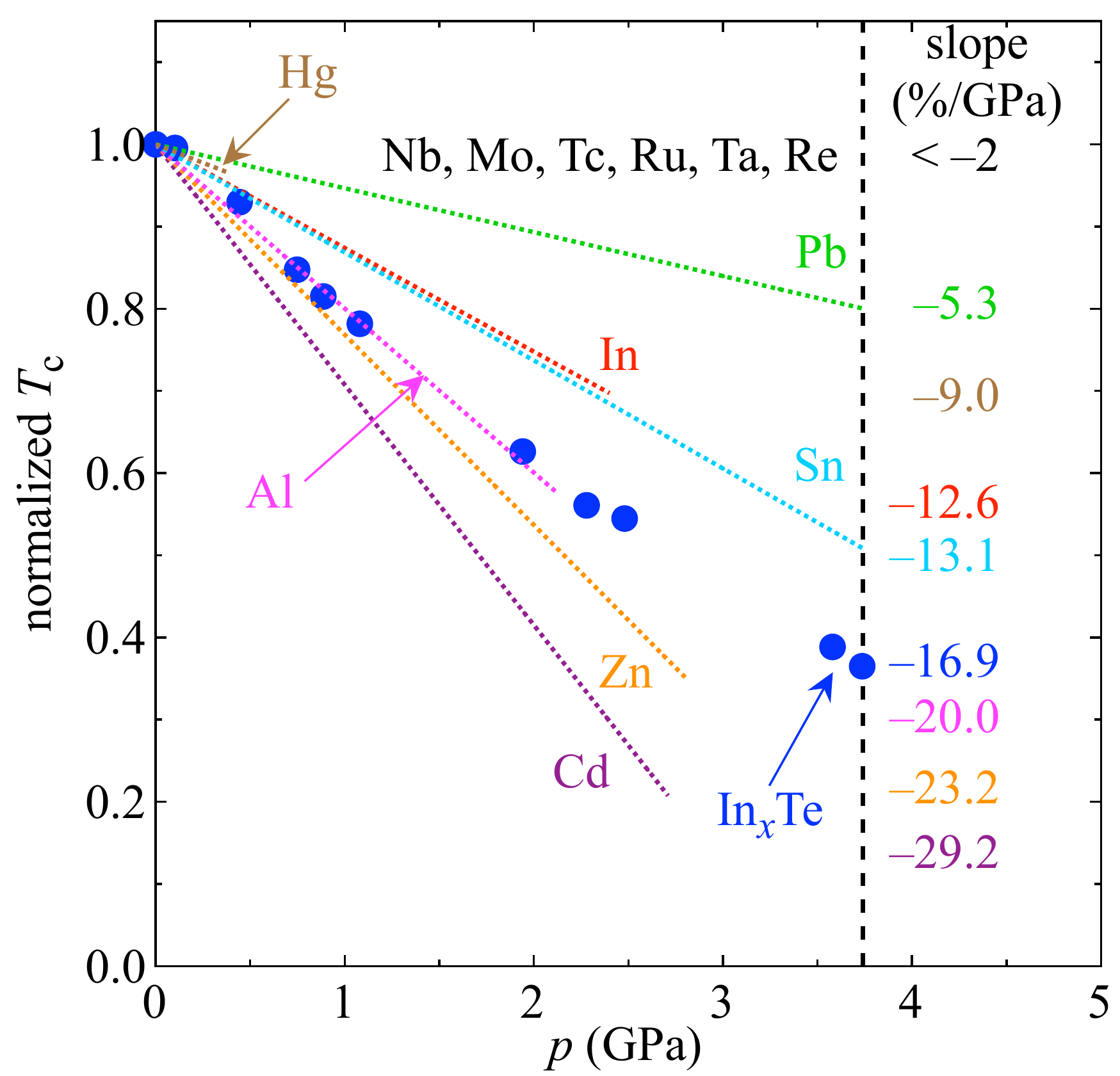}
\caption{Comparison of lattice effect on \Tc\ between \IT\ and several elemental BCS superconductors which exhibit a suppression of \Tc\ upon applying physical pressure. The numbers on the right-hand side are the slopes of the respective linear suppression of \Tc\ for each superconductor in \%/GPa. See text for details.}
\label{figS7}
\end{figure}

Figure~\ref{figS7} compares the ``negative pressure'' effect in \IT\ (i.e., expanding its cubic lattice by adding Indium) shown in Fig.~5 of the main text with the effect of physical pressure on \Tc\ in various elemental superconductors. The data of the latter were taken from literature: Al, Cd, Sn, Zn \cite{seiden69a}, In \cite{smith73a}, Hg \cite{jennings58a}, and Nb, Mo, Pb, Tc, Ru, Ta, Re \cite{boughton70a}. We note that the values of \Tc\ and $d\Tc/dp$ vary somewhat among different publications. According to Ref.~\cite{boughton70a}, Cd exhibits the strongest relative suppression effect on \Tc\ among the elements.

Figure~1(b) of the main text suggests that the expansion of the unit-cell volume $V = a_c^3$ saturates at $V^{\rm sat}\sim 6.18^3$~\AA$^3$. To convert the lattice effect in \IT\ to physical pressure $p$, the bulk modulus $B = 70$~GPa reported \cite{chattopadhyay85a} for cubic InTe is used:
\begin{equation}
p_{\rm sample} = \frac{V_{\rm sat}-V_{\rm sample}}{V_{\rm sat}}\cdot B.
\end{equation} 
The transition temperatures are normalized to the maximum $\Tc^0 \sim 3.51$~K in \IT\ ($x = 1.00$; blue circles, from magnetization measurements) and to the ambient–pressure \Tc\ values in case of the elemental superconductors. The highest pressure applied to \IT\ by reducing $x$ corresponds to $\sim 3.75$~GPa and is indicated by the vertical dashed line in Fig.~\ref{figS7}. 

For some elemental superconductors there are no literature data available up to this pressure explaining why the experimental data of these stop at lower pressure. We also note that at a certain physical pressure $\Tc(p)$ often tends to fall below the approximate linear evolution. The normalized data for elemental superconductors shown in Fig.~\ref{figS7} were fitted. The numbers given at the right-hand side of Fig.~\ref{figS7} are the corresponding slopes of the suppression of $\Tc/\Tc^0$ in \%/GPa. Apparently, the observed lattice effect in \IT\ is comparable to the largest experimentally achievable physical-pressure effects among chemically simple BCS superconductors.

\newpage
\subsection{Physical pressure effect measured on a sample In\boldmath{$_{1.00}$}Te}
\label{Tcp}

\begin{figure}[h]
\centering
\includegraphics[width=10cm,clip]{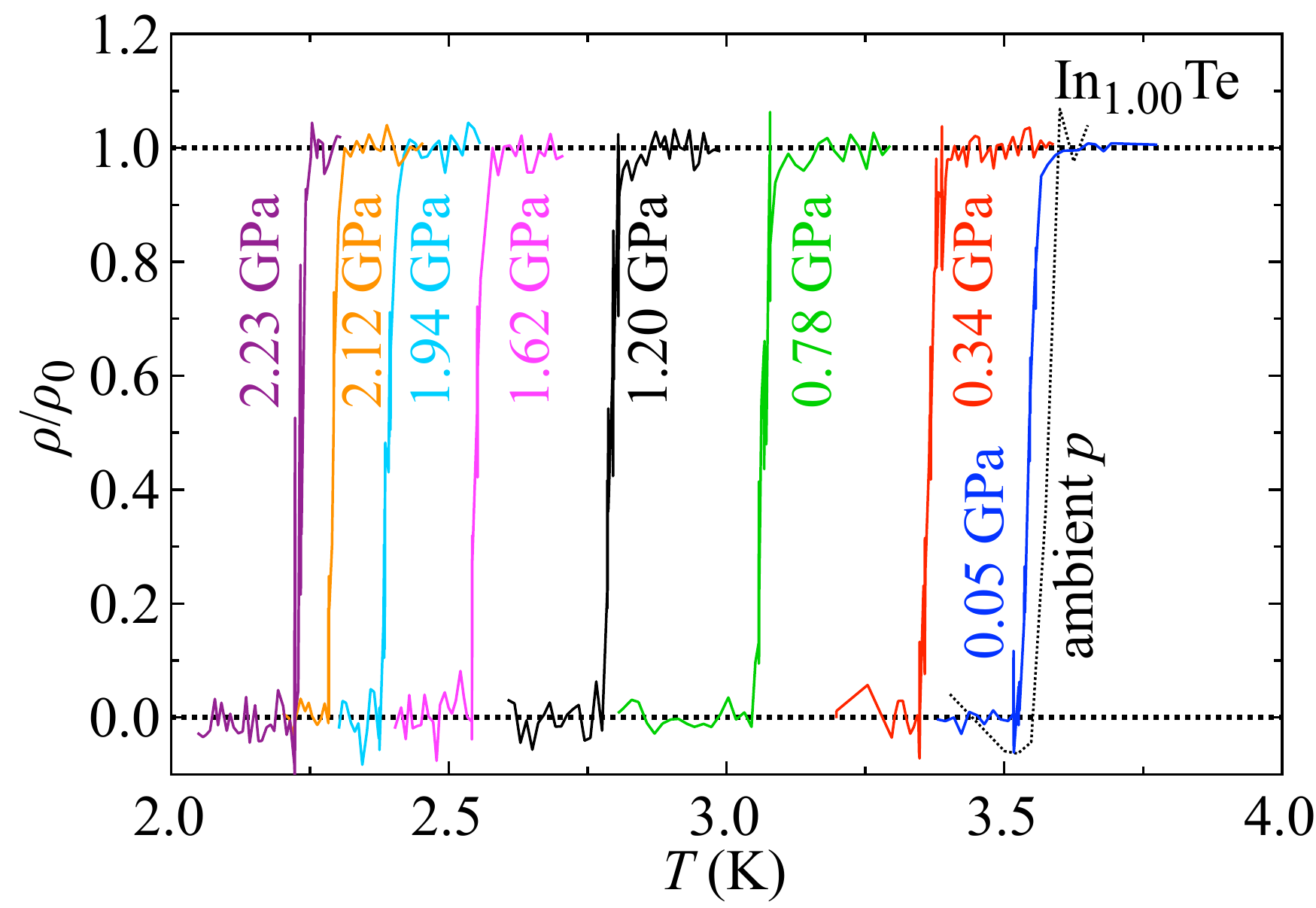}
\caption{Resistivity measurements on a stoichiometric sample In$_{1.00}$Te with an ambient-pressure $\Tc = 3.53$~K under different hydrostatic pressures normalized to the normal-state resistivity $\rho_0$.}
\label{figS8}
\end{figure}

Figure~\ref{figS8} summarizes several resistivity measurements at ambient (black dotted line) and under hydrostatic pressure $0~{\rm GPa} < p < 2.25$~GPa on a sample with $x=1.00$ and $\Tc= 3.53$~K (zero resistance) at ambient conditions. The applied physical pressure was determined from the suppression of the superconducting \Tc\ of a simultaneously measured Pb standard sample by employing published $\Tc(p)$ data for Pb: $d\Tc/dp = -0.379$~K/GPa with $\Tc=7.19$~K at ambient conditions \cite{boughton70a}. In both cases \Tc\ is defined as the temperature at which the resistance becomes zero upon decreasing temperature. Estimated data $\Tc(p)$ are plotted in Fig.~5 of the main text.

\clearpage

%

